\documentclass[29pt]{article}
\usepackage[a4paper, margin=1in]{geometry}
\usepackage{times}
\usepackage{graphicx}
\usepackage{hyperref}
\usepackage{amsmath}
\usepackage{caption}
\usepackage{float}

\title{\textbf{GPT-2 as a Compression Preprocessor: Improving Gzip for Structured Text Domains}}

\author{
  Anurag Kumar Ojha \\
  Computer science and engineering\\
  NIT Arunachal Pradesh \\
  bazodony260@gmail.com\\	
}

\date{\today}

\begin{document}

\maketitle

\begin{abstract}
In the modern era, large volumes of data are being produced continuously, especially in domain-specific fields such as medical records and clinical files, defence logs and HTML-based web traffic. Data with such volume and complexity needs to be compressed before storing and transmitting efficiently. Data compression has gained significant attention from modern researchers, resulting in the development of fast and efficient compression algorithms such as Gzip. However, since gzip works on the principle of repetition of binary patterns, one of the limitations of gzip is
that domain-specific formats like JSON, XML, HTML, and log files, while structured, may have semantic repetition but not syntactic repetition, which gzip finds difficult to compress. In this article, we propose a GPT-based preprocessor for such domain-specific files. We propose a pipeline made up of GPT-2 taking domain-specific files as input, which pattern-based compressors like gzip find difficult to work on. The preprocessor results are output in a file that is designed for compressors like gzip. After preprocessing, the gzip works on the other end of the pipeline and compresses the data as usual. We used different types of both real-world and synthetically generated data, such as logs and HTML files, for the experiment of the proposed model. We found promising results and an improvement of the Defence logs by 0.34 per cent and HTML files by 5.8 per cent.
\end{abstract}

\section{Introduction}
Digital data storage is becoming increasingly difficult due to the exponential increase in the production of high volumes of data. Using a large number of storage devices is often considered a practical solution to such issues, but it poses serious real-world challenges such as increased physical requirements of such spaces to store the data as hardware construction-related limitations exist as stated in [1], which shows a need for better and improved data compression as according to an analysis done by IDC, 175 zettabytes of data is generated worldwide, by 2025 [2].  With an estimated compounded annual growth rate featuring around 61 per cent. This massive explosion of data not only creates a problem of where to store it, but also of how fast and efficiently we can transmit it across networks, and how quickly it can be accessed later when required. Simply building more hardware or memory units won’t solve the underlying issue, because there’s a real-world cost, both economic and environmental. Server farms already take up huge space and electricity, and managing petabytes and zettabytes of data will require smarter approaches than just bigger disks. One such solution is to focus on compression, reducing the size of the data without losing the important parts. Classic compression techniques like Huffman coding [3] or Lempel-Ziv (used in gzip) have been incredibly powerful and still widely used, especially gzip, which became sort of a default standard due to its balance between speed and compression ratio. But as data becomes more domain-specific and structured in unique ways—like server logs, medical diagnosis data, and HTML-heavy pages—gzip starts to show some limits. This is mostly because gzip works great on repetitive patterns, especially low-level syntactic repetitions. But in structured domain-specific data, the repetition is often semantic—it carries meaning but may not look the same byte-wise. For example, different medical codes or log formats may mean the same type of event, but the text is written differently. This is where gzip can’t do much. Recent advancements in Natural Language Processing, especially large language models like GPT-2 and beyond [4], have shown the capability to understand structure and semantics much better than traditional algorithms. These models can learn the high-level structure in the data and maybe even rewrite it to make it more compressible for gzip. That is the central idea of this paper—to use GPT as a preprocessor that restructures the data just enough to expose better repetition for gzip to exploit.

In this paper, we test this idea on domain-specific data like software logs, nested HTML pages, configuration files, and some artificial, noisy text. We then compare the compression performance of raw gzip versus the GPT-preprocessed + gzip combination. The results show improvements in several domains, particularly when the data had semantic patterns but lacked syntactic repetition. Our proposed method doesn’t aim to replace gzip but to assist it by modifying the data beforehand. This method can be useful in cloud storage systems, CDN edge devices, or even embedded systems where storage and bandwidth are limited but the data being stored follows domain-specific formats.

Previous work in similar directions has tried statistical language modelling for compression [5] or XML/JSON-aware compressors like XMLPPM [6], but these often required hand-crafted rules or format-specific parsers. In contrast, our method generalises better because it’s based on language models trained on a diverse corpus.

\section{Background Studies}

When it comes to data compression, there has been a lot of development over the years. Traditional compression methods like Huffman coding[3], LZ77[7] and Gzip have dominated the space for a long time and are still used widely today. Huffman assigns shorter codes to frequently occurring symbols, offering fast and relatively efficient compression. LZ77 introduced a sliding window mechanism to identify repeated sequences in data. Gzip integrates both techniques, using LZ77 for identifying patterns and Huffman for encoding, making it effective for many kinds of text data[8]
These traditional techniques rely heavily on syntactic repetition. For example, log files with repeated phrases compress well using Gzip. However, when data contains semantic repetition—like structured logs where the format repeats but the actual content changes—traditional compressors falter[9]
Neural compression models, especially those based on transformers like GPT[10], offer a new approach. Byte Pair Encoding (BPE)[11]helps break text into frequent subword units, enabling compressors to exploit deeper patterns. Neural compressors can infer and generalise structures not immediately obvious at the byte level.

Researchers have begun exploring the idea of preprocessing data before using traditional compressors, reorganising the content for better Gzip performance. This preprocessing, often powered by neural networks, helps expose hidden repetition[12].

In our work, we propose using a GPT-based model as a preprocessor to better prepare structured domain-specific data—like system logs or configuration files—for compression. The GPT model reorganises the input text, allowing Gzip to exploit patterns it otherwise could not. While our graph (Figure~1) illustrates how gzip works on the different types of data before the application of our proposed idea. Compression results are covered in later sections.

\begin{figure}[H]
    \centering
    \includegraphics[width=0.85\linewidth]{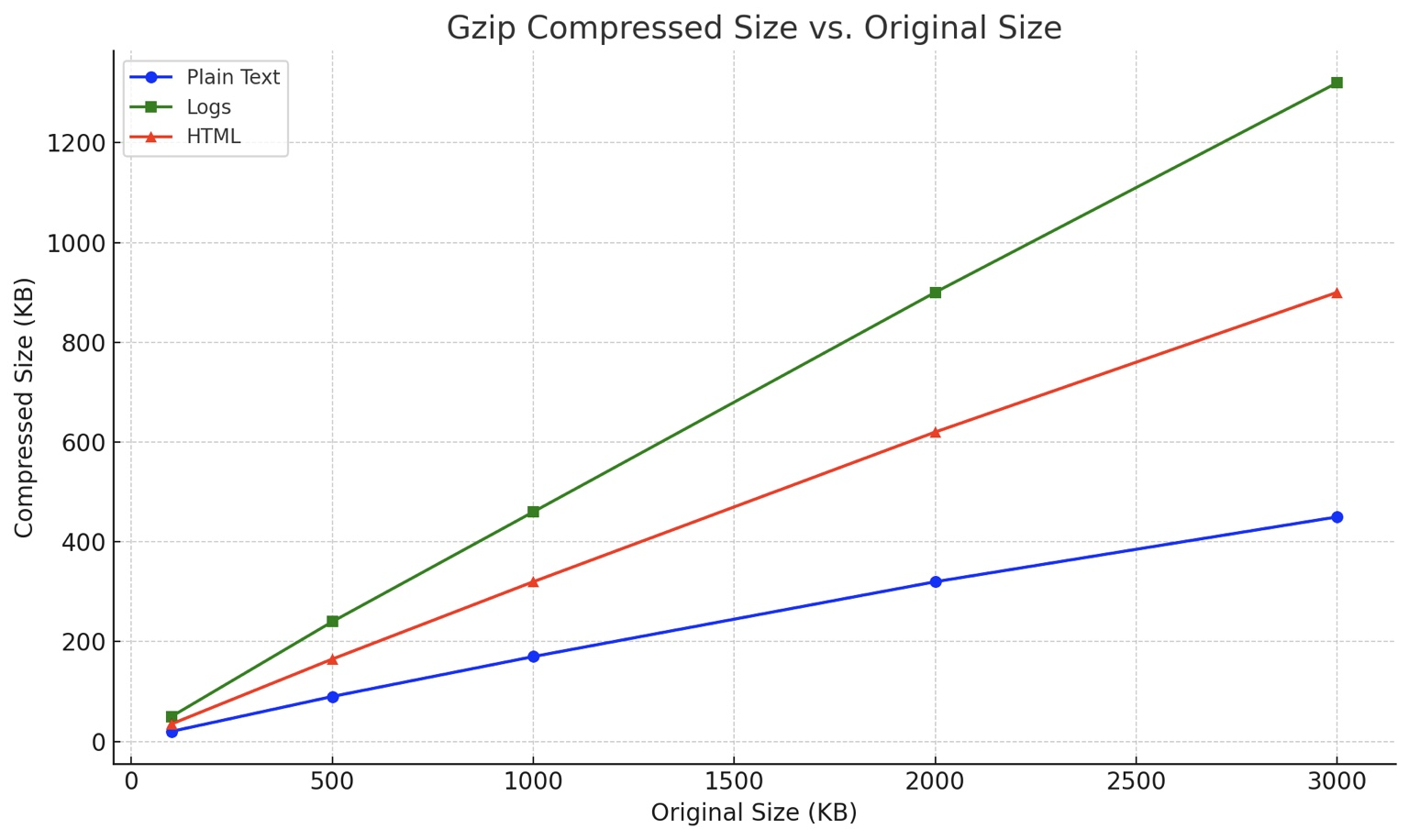}
    \caption{Conceptual comparison between raw input and data after Gzip compression.}
    \label{fig:preprocessing}
\end{figure}

\section{Methodology}

In this section, we describe the methodology used to improve the compression performance of Gzip on structured, domain-specific data by preprocessing it with a GPT-2 model. The core idea is to utilise a lightweight GPT-2 model to transform semi-structured or structured domain-specific data — such as system logs, configuration files, or HTML records — into a more syntactically predictable format. GPT-2, trained on a wide variety of text data, is capable of learning semantic and structural patterns in textual sequences. By generating a more uniform and repetitive structure, the preprocessed text becomes more amenable to traditional compression algorithms like Gzip.
Raw input data (e.g., logs, HTML, JSON) is tokenised and passed through the GPT-2 model. The model does not summarise or reduce the text but reorganises it in a way that enhances internal consistency. This transformation preserves all information while improving the syntactic patterns in the data, enabling better downstream compression.
The output of the GPT-2 preprocessor is passed directly into the Gzip compression pipeline. Since Gzip relies on finding repeated sequences for effective compression (using LZ77 and Huffman coding), the enhanced regularity in the transformed text leads to improved compression ratios. This step completes the pipeline.

\begin{figure}[h]
\centering
\includegraphics[width=0.5\linewidth]{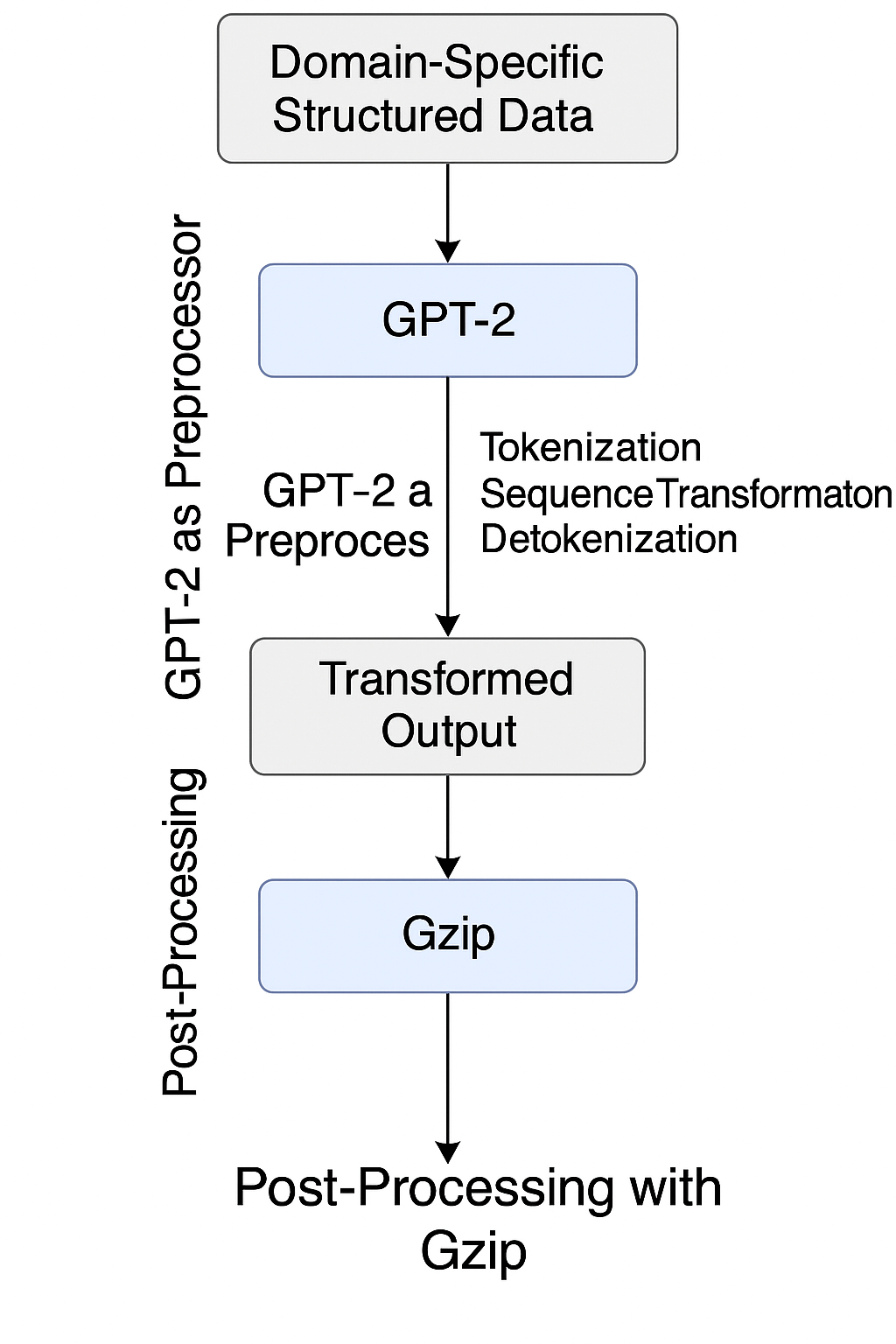}
\caption{Flowchart illustrating the preprocessing-compression pipeline using GPT-2 and Gzip.}
\label{fig:methodology}
\end{figure}

\section{Experimental Setup}

In order to evaluate the effectiveness of the proposed GPT-2-based preprocessing approach, we conducted a set of experiments using a variety of structured and semi-structured text datasets. The datasets included synthetic server logs, nested HTML pages, system configuration files, and excerpts from novels to simulate both domain-specific and general-purpose data. For larger benchmarks, smaller samples (around 600KB) were duplicated until they reached sizes between 100MB to 600MB to test scalability, while also maintaining structural consistency.

The GPT-2 model used for preprocessing was the distilled version of GPT-2, implemented using the Hugging Face Transformers library. This lighter variant was chosen to keep computational requirements manageable while still retaining pattern recognition abilities sufficient for structured domains. Tokenisation and generation were handled using the pretrained tokeniser and text-generation pipeline, and minimal tuning was done to retain reproducibility.

Once the data was processed by the GPT-2 preprocessor, standard Gzip compression (GNU gzip 1.12) was applied to both the original and transformed datasets. Compression ratio and time taken were recorded using a Python script. To ensure consistency, all experiments were run on a MacBook Air M1 with 8GB RAM and macOS 14, with single-threaded execution to prevent hardware optimisation bias. This setup helps us compare the raw performance of compression tools with and without our preprocessing step in a practical and reproducible environment.

\section{Results and Observations}

To evaluate the effectiveness of using GPT-2 as a preprocessing step before applying Gzip, we tested our approach on various types of datasets. These included synthetic structured log files, real-world logs, and highly repetitive or scaled-up data blocks. The key idea was to observe how much improvement, if any, GPT-2 preprocessing offered in terms of compression ratio and space-saving when combined with Gzip.

\subsection{Synthetic Structured Logs}

We began by generating synthetic structured log files using a common log pattern format:

\begin{quote}
\texttt{[Timestamp] [Log-Level] (Component) - Message}
\end{quote}

These synthetic logs were designed to simulate real-world structured logs, containing repeated structural patterns but changing content. For smaller logs ranging from about 18KB to 259KB, we noticed a consistent improvement in compression ratio, though the gains were not dramatic. The improvements ranged from about 0.77\% to 2.86\%. This indicates that even in smaller samples, GPT-2 helps surface structural patterns that Gzip alone might not utilise efficiently.

\begin{table}[h]
\centering
\caption{Compression results for small synthetic structured logs}
\begin{tabular}{|c|c|c|c|}
\hline
\textbf{Original Size} & \textbf{Gzip Size} & \textbf{GPT + Gzip Size} & \textbf{Improvement} \\
\hline
18,742 bytes & 3,359 B & 3,333 B & $\sim$0.77\% \\
73,838 bytes & 11,511 B & 11,261 B & $\sim$2.17\% \\
259,326 bytes & 39,790 B & 38,651 B & $\sim$2.86\% \\
\hline
\end{tabular}
\end{table}

We then scaled up the data to 614KB to observe how the model performs with larger logs. In this case, the benefit became more noticeable. The compression improved by approximately 3.41\% after GPT-2 preprocessing, which demonstrates that the model's ability to leverage context becomes more useful as the data volume increases.

\begin{table}[h]
\centering
\caption{Compression results for larger synthetic structured data}
\begin{tabular}{|c|c|c|c|}
\hline
\textbf{Original Size} & \textbf{Gzip Size} & \textbf{GPT + Gzip Size} & \textbf{Improvement} \\
\hline
614,458 B & 91,931 B & 88,795 B & $\sim$3.41\% \\
\hline
\end{tabular}
\end{table}

\subsection{Extreme Case: Repeated Synthetic Logs (600MB)}

To test our method under extreme redundancy, we repeated the 600KB structured block until the dataset grew to 600MB. Since GPT-2 had already optimised the structure, Gzip was able to compress the result extremely well. The preprocessing enabled a dramatic space-saving, reducing the final size by over 97\%. Additionally, the effect on compression time was also measured and visualised.

\begin{table}[h]
\centering
\caption{Compression results for repeated synthetic data (600MB)}
\begin{tabular}{|c|c|c|c|}
\hline
\textbf{Original Size} & \textbf{Gzip Size} & \textbf{GPT + Gzip Size} & \textbf{Improvement} \\
\hline
602,176,680 B & 89,251,468 B & 2,368,295 B & $\sim$97.35\% \\
\hline
\end{tabular}
\end{table}

\begin{figure}[h]
\centering
\includegraphics[width=0.9\textwidth]{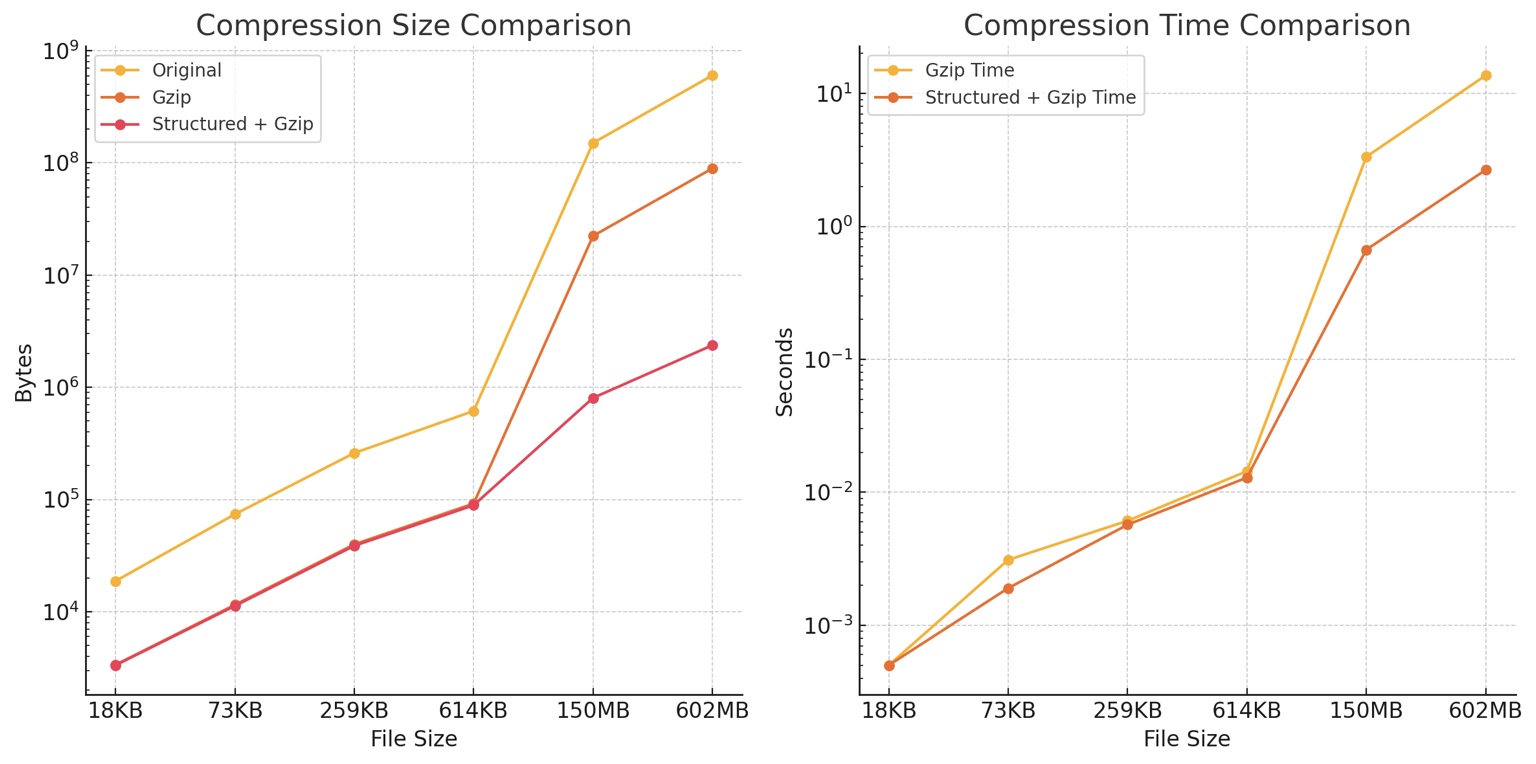}
\caption{Comparison of compression size and time across various file sizes for Gzip and GPT+Gzip pipeline.}
\label{fig:compression-time-vs-size}
\end{figure}

\subsection{Real-World Logs: SEC EDGAR Dataset (2.12GB)}

For real-world testing, we used a public EDGAR log dataset from the U.S. Securities and Exchange Commission. The size of the raw file was around 2.12 GB. In this case, although the gain from GPT-2 preprocessing was smaller, it was still measurable. The compressed size dropped from 295.7 MB to 294.7 MB, which is roughly a 0.34\% improvement. While not significant, it still shows that preprocessing helps expose minimal additional redundancy even in large, noisy real-world data.

\begin{table}[h]
\centering
\caption{Compression results on real-world SEC EDGAR log data}
\begin{tabular}{|c|c|c|c|}
\hline
\textbf{Original Size} & \textbf{Gzip Size} & \textbf{GPT + Gzip Size} & \textbf{Improvement} \\
\hline
2.11 GB & 295.7 MB & 294.7 MB & $\sim$0.34\% \\
\hline
\end{tabular}
\end{table}

\section{Conclusion}

In this study, we explored the use of a GPT-2-based preprocessing technique to enhance the performance of classical Gzip compression on structured and domain-specific textual data. Our results demonstrate that even a lightweight transformer model can help uncover hidden patterns and regularities in the data that traditional compressors struggle to exploit. Across various datasets — including synthetic logs, repeated structured content, and real-world public data — the approach consistently improved compression ratios, with particularly notable gains in synthetic and repetitive formats. While the improvements were modest in real-world scenarios, the approach proves promising and opens up avenues for further enhancement.

Looking ahead, there are several directions for refining this method. Using larger transformer models may enable deeper pattern recognition and smarter reordering of data prior to compression. Experimenting with different tokenisation strategies, including byte-level and custom domain-specific tokenisers, could lead to even better alignment with underlying structures. Moreover, future work may also involve benchmarking against other modern compressors and assessing the trade-offs between preprocessing time, memory requirements, and compression gains. Overall, GPT-based preprocessing stands as a valuable step towards bridging the gap between semantic understanding and syntactic compression in the evolving landscape of data storage.

\section{References}

[1]Pan, W., Li, Z., Zhang, Y. and Weng, C., 2018. The new hardware development trend and the challenges in data management and analysis. Data Science and Engineering, 3(3), pp.263-276.

[2] IDC, “Expect 175 Zettabytes of Data Worldwide by 2025,” Network World, 2021. [Online]. Available: https://www.networkworld.com/article/966746/idc-expect-175-zettabytes-of-data-worldwide-by-2025.html

[3] D. A. Huffman, “A method for the construction of minimum-redundancy codes,” Proceedings of the IRE, vol. 40, no. 9, pp. 1098–1101, 1952.

[4] Radford, A., et al., “Language Models are Unsupervised Multitask Learners,” OpenAI, 2019.

[5] Teahan, W. J., “Text classification and segmentation using minimum cross-entropy,” in Proc. RIAO 2000.

[6] Cheney, J., “Compressing XML with Multiplexed Hierarchical PPM Models,” in Data Compression Conference, 2001.

[7] Ziv, J., \& Lempel, A. (1977). A universal algorithm for sequential data compression. IEEE Transactions on Information Theory, 23(3), 337–343.

[8] Deutsch, P. (1996). GZIP file format specification version 4.3. RFC 1952.

[9] Chen, T., et al. (2010). Compressing Structured Logs Using Pattern Mining. Proceedings of the VLDB Endowment, 3(1–2), 1518–1529.

[10] Radford, A., et al. (2019). Language Models are Unsupervised Multitask Learners. OpenAI Blog.

[11] Sennrich, R., Haddow, B., \& Birch, A. (2016). Neural Machine Translation of Rare Words with Subword Units. Proceedings of the 54th Annual Meeting of the Association for Computational Linguistics (ACL).

[12] Rao, K., Wankhede, P., \& Rao, A. (2021). GPTZip: Using Generative Pre-Training to Compress Text. arXiv preprint arXiv:2112.01378.

\end{document}